\documentclass[sigconf]{acmart}

\usepackage{multirow}
\usepackage{graphicx}
\usepackage{dblfloatfix}

\AtBeginDocument{%
  }

\copyrightyear{2024}
\acmYear{2024}
\setcopyright{acmlicensed}\acmConference[UbiComp Companion '24]{Companion of the 2024 ACM International Joint Conference on Pervasive and Ubiquitous Computing Pervasive and Ubiquitous Computing}{October 5--9, 2024}{Melbourne, VIC, Australia}
\acmBooktitle{Companion of the 2024 ACM International Joint Conference on Pervasive and Ubiquitous Computing Pervasive and Ubiquitous Computing (UbiComp Companion '24), October 5--9, 2024, Melbourne, VIC, Australia}
\acmDOI{10.1145/3675094.3678453}
\acmISBN{979-8-4007-1058-2/24/10}

\begin{document}

\title{Left-Right Swapping and Upper-Lower Limb Pairing for Robust Multi-Wearable Workout Activity Detection}

\author{Jonas Van Der Donckt}
\authornote{Both authors contributed equally to this research.}
\orcid{1234-5678-9012}
\affiliation{%
  \institution{IDLab, Ghent University-imec}
  \city{Ghent} \country{Belgium}
}
\email{jonvdrdo.vanderdonckt@ugent.be}

\author{Jeroen Van Der Donckt}
\authornotemark[1]
\affiliation{%
  \institution{IDLab, Ghent University-imec}
  \city{Ghent} \country{Belgium}
}
\email{jeroen.vanderdonckt@ugent.be}

\author{Sofie Van Hoecke}
\affiliation{%
  \institution{IDLab, Ghent University-imec}
  \city{Ghent} \country{Belgium}}
\email{sofie.vanhoecke@ugent.be}

\renewcommand{\shortauthors}{Van Der Donckt et al.}

\begin{abstract}
This work presents the solution of the Signal Sleuths team for the 2024 HASCA WEAR challenge. The challenge focuses on detecting 18 workout activities (and the null class) using accelerometer data from 4 wearables -- one worn on each limb. Data analysis revealed inconsistencies in wearable orientation within and across participants, leading to exploring novel multi-wearable data augmentation techniques. We investigate three models using a fixed feature set: (i) "raw": using all data as is, (ii) "left-right swapping": augmenting data by swapping left and right limb pairs, and (iii) "upper-lower limb paring": stacking data by using upper-lower limb pair combinations (2 wearables). Our experiments utilize traditional machine learning with multi-window feature extraction and temporal smoothing. Using 3-fold cross-validation, the raw model achieves a macro F1-score of 90.01\%, whereas left-right swapping and upper-lower limb paring improve the scores to 91.30\% and 91.87\% respectively.
\end{abstract}


\begin{teaserfigure}
  \includegraphics[width=\textwidth]{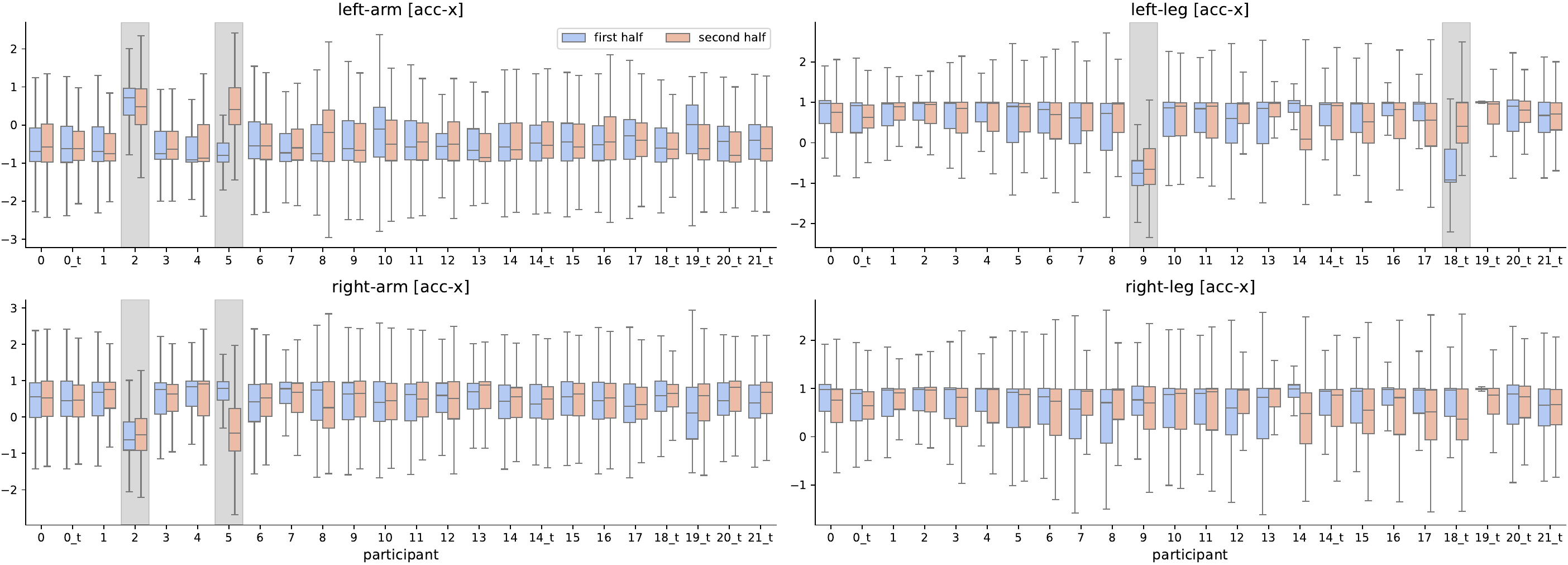}
  \caption{Distribution of x-axis accelerometer data for each participant recording, split into first and second halves, for all four limbs. Deviating patterns are highlighted in gray, \textit{\_t} indicates whether the participant recording belongs to the test data.}
  \Description{Four subplots are shown, showing the distribution of the x-axis accelerometer data for each participant recording via a box plot for each limb.}
  \label{fig:teaser}
\end{teaserfigure}

\received{4 July 2024}
\received[revised]{12 July 2024}
\received[accepted]{26 July 2024}

\maketitle

\section{Introduction}
Data-driven processing of wearable data facilitates numerous applications, ranging from automatic activity recognition~\cite{wearables_for_human_2015,stojchevska2024unlocking}, aiding in tracking workout progress or gathering long-term health insights~\cite{van2022self}, to intelligent service adaptations, such as adaptive heart-rate monitoring based on detected activities~\cite{yan2012energy}. 

Recently, Bock et al. introduced the WEAR dataset~\cite{bock2023wear}, which utilizes an acquisition protocol involving four wearable devices (one worn on each limb) collecting three-axial accelerometer data and includes both first- and third-person video data. The dataset consists of 18 participants, each performing 18 workout activities in one or multiple sessions. 
High-quality activity labels for the 18 workout activities and a "null" class (thus 19 classes in total) were obtained by using the third-person video data. The WEAR dataset stands out due to its combination of both multi-wearable and video data, and the availability of untrimmed, continuous data streams for each session. 

Given our track record with multimodal time series classification using traditional machine learning, we opted to not employ deep learning for this challenge~\cite{sleep_linear_model}. Our proposed pipeline consists of (i) multi-window feature extraction, (ii) data augmentation, (iii) a traditional machine learning model (i.e., CatBoost - gradient boosted trees), and (iv) post-processing of model predictions. In particular, we design and evaluate two novel data augmentation techniques for multi-wearable data; \textit{left-right swapping} and \textit{upper-lower limb pairing}. As such, we aim to construct multiple robust models whose predictions are leveraged during post-processing to further enhance the performance.


\section{2024 WEAR Dataset Challenge}
The 2024 HASCA WEAR challenge aims to detect 18 distinct workout activities (and a null class), using inertial data from multiple wearable devices. Each participant wore a Bangle.js v1 watch~\cite{espruino_bangle} on every limb (left-arm, right-arm, left-leg, right-leg) in a fixed orientation. All four wearables collected three-axial accelerometer data at 50 Hz, with a range of [-8g, 8g]. Participants were suggested to perform 18 different workout activities for $\pm$90 seconds and in two sessions, i.e., $\pm$ 9 activities per session, which were combined into a single continuous recording.

The initial dataset, described in the WEAR dataset paper~\cite{bock2023wear}, includes data from 18 participants and serves as the training set for this challenge. The test set contains recordings of 6 participants; of which two participants are also present in the training set.
The challenge is evaluated using the sample-wise macro F1 score. Notably, the WEAR dataset paper~\cite{bock2023wear} provides inertial baselines along with their validation procedures, allowing challenge participants to position the performance of their approaches.

A particular interesting aspect of the WEAR challenge is that the data is provided "as-is" in its untrimmed format, without any segmentation. As such, challenge participants have full flexibility to define their data windows, strides, and post-processing rules. 

Although each participant performed the same set of 18 workout activities, they had the freedom to determine the exact activity sequence and how each sequence was performed (multiple short sessions or one large session). This variability makes the dataset more representative of real-life workout scenarios. Additionally, several workout activities are similar to, or variants of others (e.g., lunges vs. lunges complex), adding a significant layer of complexity to the challenge.

\subsection{Exploratory Analysis}
\begin{figure}
    \centering
    \includegraphics[width=\linewidth]{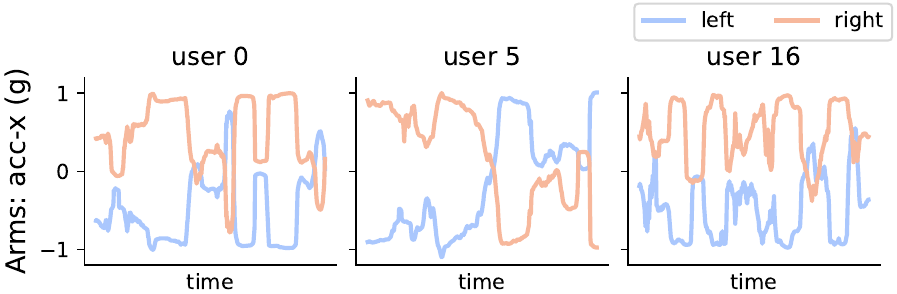}
    \vspace{-5mm}
    \caption{Median-smoothed arm accelerometer x-axis data using a 120-second smoothing window for three participants from the training set.}
    \Description{Three horizontal stacked subplots are shown, indicating the average acc-x value for train set user 0, 5, and 16, respectively.}
    \label{fig:orientation_swap}
\end{figure}

In the WEAR dataset paper, the authors stated that the wearables were worn in a fixed orientation~\cite{bock2023wear}. However, our exploratory analysis, outlined below, revealed several inconsistencies in device orientation both across- and within-participant recordings. 

Teaser Figure~\ref{fig:teaser} displays the distribution of x-axis accelerometer data for each participant across all four wearables. We focused on the x-axis accelerometer data because, as shown in Figure 1 of the WEAR dataset paper~\cite{bock2023wear}, the x-axis should align with the gravity component (i.e., 1 g), with the sign depending on the wearable orientation. In Figure~\ref{fig:teaser}, we split the data for each participant into the first and second halves, roughly corresponding to the two suggested recording sessions. 

From Figure~\ref{fig:teaser}, we observe that for participant 2, the orientation of the left- and right-arm wearables appear to be inverted in both sessions. Similarly, the left-leg data for participant 9 seems to be inverted in the two sessions. This suggests that participant 9 likely wore the wearable in the opposite orientation, while for participant 2, it is also plausible that the left- and right-arm wearables were swapped.

Additionally, participant 5 and test participant 18 exhibit changes in orientation between their sessions (i.e., between first and second half). We suspect that test participant 18 wore the wearable in opposite orientation during the first session. For participant 5, it is possible that the left- and right-arm wearables were worn on incorrect arms during the second session, or that they were worn on the correct arm but in opposite orientation. Figure~\ref{fig:orientation_swap} complements the observation for participant 5 by showing the median-smoothed x-axis accelerometer signal from the wearables on both the left and right arms. The switch in the median value midway for participant 5 indicates a change in device orientation. Such within-participant changes in orientation likely result from participants accidentally switching the wearable orientation between sessions, as all sessions are combined into a single user file. We hypothesize that dealing with these variations in orientation, as opposed to incorrectly assuming a fixed orientation, will lead to the design of more robust activity detection models.

Finally, we observed that for user 10, a large segment of left-arm data was missing. This missing data segment was mitigated by imputing the missing segment with the right-arm data. 
Also for test set participant 19, Figure~\ref{fig:teaser} indicates that for the first session, there is nearly no variation in the x-axis data for both legs, contrary to his second session recording and all other participants.

\section{Algorithm Pipeline}

\subsection{Preprocessing}
In addition to the raw data, we will also consider the Signal Magnitude Vector (SMV) for feature extraction. To do so, the SMV is calculated in a preprocessing step for each wearable $\mathcal{W}$ using the formula:
$$SMV_{t,\mathcal{W}} = \sqrt{\mathcal{W} _{t,x}^2 + \mathcal{W} _{t,y}^2 + \mathcal{W} _{t,z}^2}$$


\subsection{Feature Extraction}

\begin{table}[t]
\caption{Overview of computed features (N=14).}
\vspace{-3mm}
\label{tab:features}
\resizebox{\columnwidth}{!}{%
\begin{tabular}{llll}
\toprule
\multicolumn{1}{c}{\textbf{Domain}} & \multicolumn{1}{c}{\textbf{Notes}} & \multicolumn{1}{c}{\textbf{Features}} & \multicolumn{1}{c}{\textbf{n}} \\
\midrule
 & PSD & Spectral entropy & 1 \\ 
 \hline
\multirow{7}{*}{\textbf{Time}} & \multirow{6}{*}{RAW / SMV signals} & min, max, ptp, iqr & 4 \\
 &  & std, skew, kurtosis & 3 \\
 &  & Hjorth mobility \& complexity & 2 \\
 &  & mean crossing rate & 1 \\
 & & differential entropy & 1 \\
 & & Petrosian fractal dimension & 1 \\
 & & Katz fractal dimension & 1 \\
 \bottomrule
\end{tabular}%
}
\end{table}

\begin{figure}[t]
    \centering
    \includegraphics[width=\linewidth]{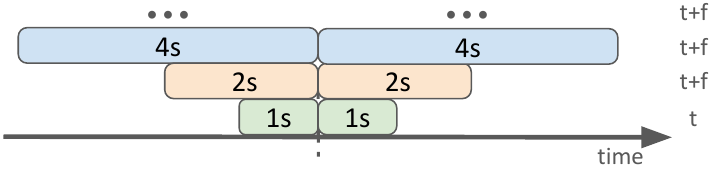}
    \vspace{-6mm}
    \caption{Multi-resolution feature extraction. For each prediction timestamp, future and past window sizes of \{ 1s, 2s, 4s, 8s, 16s, 32s \} are used, resulting in 6 $\times$ 2 feature windows. For the 1-second window, only time domain features (\texttt{t}) are extracted, while for all larger windows, both time and frequency domain features (\texttt{t+f}) are computed.}
    \Description{Visual illustration of multi-resolution feature extractions.}
    \label{fig:multi-res-features}
\end{figure}

To construct a performant feature vector, we relied on multi-reso\-lution feature extraction, a method proven to be highly effective for continuous time-series classification and capable of matching state-of-the-art deep learning approaches~\cite{sleep_linear_model}. 
Figure~\ref{fig:multi-res-features} illustrates our utilized multi-resolution window configuration for each prediction time-step. Feature windows that start and end at the prediction time-step are incorporated in the feature vector. We use a 0.5 second stride, meaning each consecutive time-step has a 0.5 second gap. 

Table~\ref{tab:features} provides an overview of the extracted features on each window. Our \texttt{tsflex} toolkit was used for convenient and efficient multi-window feature extraction~\cite{van_der_donckt_tsflex_2022}. 
Spectral entropy, fractal, and Hjorth features were computed using the \texttt{antropy} toolkit~\cite{vallat_antropy}, while other features were derived from functions provided by the \texttt{numpy} and \texttt{scipy} libraries~\cite{harris_array_2020, virtanen2020scipy}. 

In total, 14 features were extracted per window, resulting in 166 multi-window features per wearable accelerometer axis for each time-step: 2 (start and end) $\times$ 6 windows $\times$ 14 features - 2 (spectral entropy is not computed on 1-second windows). Overall, this results in 1992 features; 166 features x 3 accelerometer axes x 4 wearables.


\subsection{Data augmentation}
In this study, we propose and investigate three data augmentation techniques, aimed at creating more robust models. The first technique, \textit{rotation-invariant} statistical aggregation, is a generic approach that aims to effectively discard axial information while retaining the overall feature information. The other two proposed techniques, \textit{left-right swapping} and \textit{upper-lower limb pairing}, expand the data and are solely applicable to multi-wearable setups.  

\subsubsection{Rotation-invariant aggregation}
To improve the robustness of feature sets computed on raw axial signals (i.e., non-SMV transformed), we examined three novel \textit{rotation-invariant statistical aggregation} methods, which are applicable to any three-axial modality, such as gyroscope, magnetometer or here accelerometer data. Specifically, the raw $\{x, y, z\}$ signal features are condensed into summary statistics using: \texttt{stat2}: $\{$ mean, std $\}$, \texttt{stat3}: $\{$ mean, std, skew $\}$, or \texttt{sort}: $\{$ min, mid, max $\}$. Similar to SMV, these methods discard axial information while retaining overall feature information. Notably, \texttt{stat2} compresses the three ${x, y, z}$ features into two summary statistics (a 1/3 compression ratio), while \texttt{stat3} and \texttt{sort} maintain the same input-output ratio.

\subsubsection{Left-Right Swapping}
Left-Right swapping (LR-swapping) is a multi-wearable data augmentation technique that expands the feature matrix by including all possible combinations of upper and lower left-right swaps. Specifically, the data is augmented with the following combinations: \{"no-swap", "upper LR-swap", "lower LR-swap", "upper \& lower LR-swap" \}. Remark how the "no-swap" group corresponds to the feature matrix used in the "default" (i.e., raw) configuration.
Remark that this swapping can also be performed on the validation data, which can then be aggregated by using the majority voting across the different swap combinations for the same prediction time-step.

\subsubsection{Upper-Lower Limb Paring}
Upper-Lower limb pairing (UL-pairing) is a multi-wearable augmentation technique that pairs one upper limb with one lower limb per feature vector, resulting in the use of two wearables per feature vector instead of four. The data is augmented by forming all possible upper-lower limb pair combinations: \{ left-arm \& left-leg, left-arm \& right-leg, right-arm \& left-leg, right-arm \& right-leg \}. This approach increases the original feature matrix size by 4$\times$ but reduces the number of features (in the vector) by half. 
Similar to LR-swapping, this technique can be applied on the validation data, followed by prediction time-step aggregation of the different combinations.

\subsection{Model}

Given the power of traditional machine learning models matching deep learning in certain time-series classification tasks~\cite{sleep_linear_model}, we focused on traditional machine learning models, specifically evaluating the performance of CatBoost. Catboost is a gradient-boosted trees algorithm known for its strong performance without requiring extensive parameter tuning~\cite{prokhorenkova2018catboost}, making it particularly suitable for conducting an ablation study on the above-introduced data augmentation techniques. 
For our experiments, we limited the CatBoost model to 1,000 iterations (or trees) and set the \texttt{auto\_class\_weight} parameter to "Balanced". To speed up the training, we restricted the tree depth to 5 and used a border count of 32. Other parameters were kept to their default values.

\begin{figure*}[b]
    \centering
    \includegraphics[width=\linewidth]{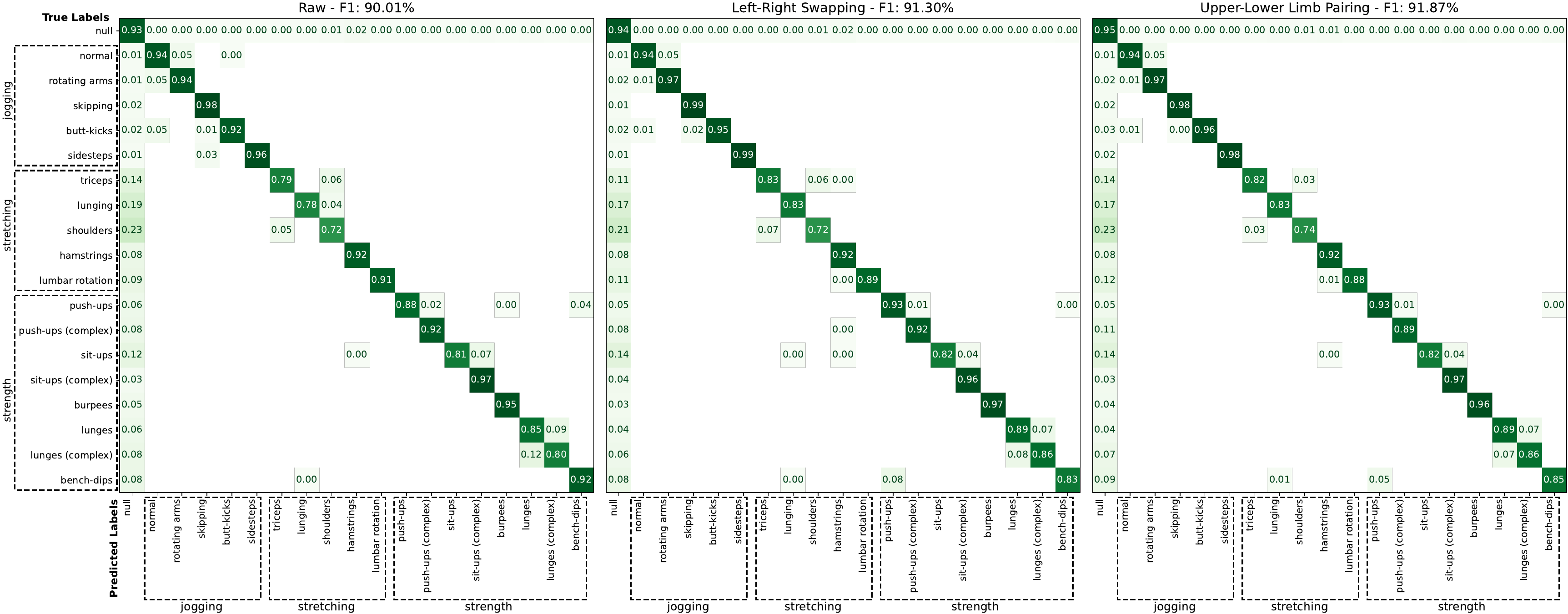}
    \vspace{-5mm}
    \caption{Normalized confusion matrix of the grouped 3-fold predictions (with temporal prediction smoothing).}
    \Description{Three confusion matrices are shown, representing the predictions of the raw configuration, LR-swapping, and UL-pairing configuration.}
    \label{fig:cm-3-fold}
    \vspace{-3mm}
\end{figure*}

\subsection{Postprocessing}\label{sec:postprocess}
To improve our predictions, we applied two commonly used postprocessing techniques: \textit{k-fold majority voting}, which exploits model diversity, and \textit{temporal prediction smoothing}, which leverages the high temporal correlation of labels.
As a final boost to our test set predictions, we performed \textit{rule-based activity boosting}, which exploits the characteristics of the study protocol, making this boost not applicable in practice.

\subsubsection{K-fold Majority Voting}
Instead of training on the entire train dataset to make final predictions on the test data, we utilized K-fold cross-validation (CV). The predictions made after fitting each fold are aggregated by utilizing majority voting (MV) on the prediction probabilities. We deliberately opted for a 3-fold CV, as this results in only 50\% data overlap across the training folds. Using this approach, we effectively perform bootstrapping through CV on the training data. When using an empirical leave-one-subject-out (LOSO) 3-Fold CV approach, we observed that this majority voting aided in boosting the performance with 1\% (absolute) on average. 

\subsubsection{Temporal Prediction Smoothing} 
We implemented prediction smoothing to leverage the high temporal correlation between consecutive feature vectors (and thus labels). This was done using a vectorized approach with a distribution function to weight the smoothing process. Specifically, for predictions with a 0.5s stride, we utilized a 10-10 receptive left-right field (i.e., 5s on each side) and a normal distribution function with $\sigma=6$.

\subsubsection{Rule-based activity boosting}
\begin{figure*}[b]
  \includegraphics[width=\textwidth]{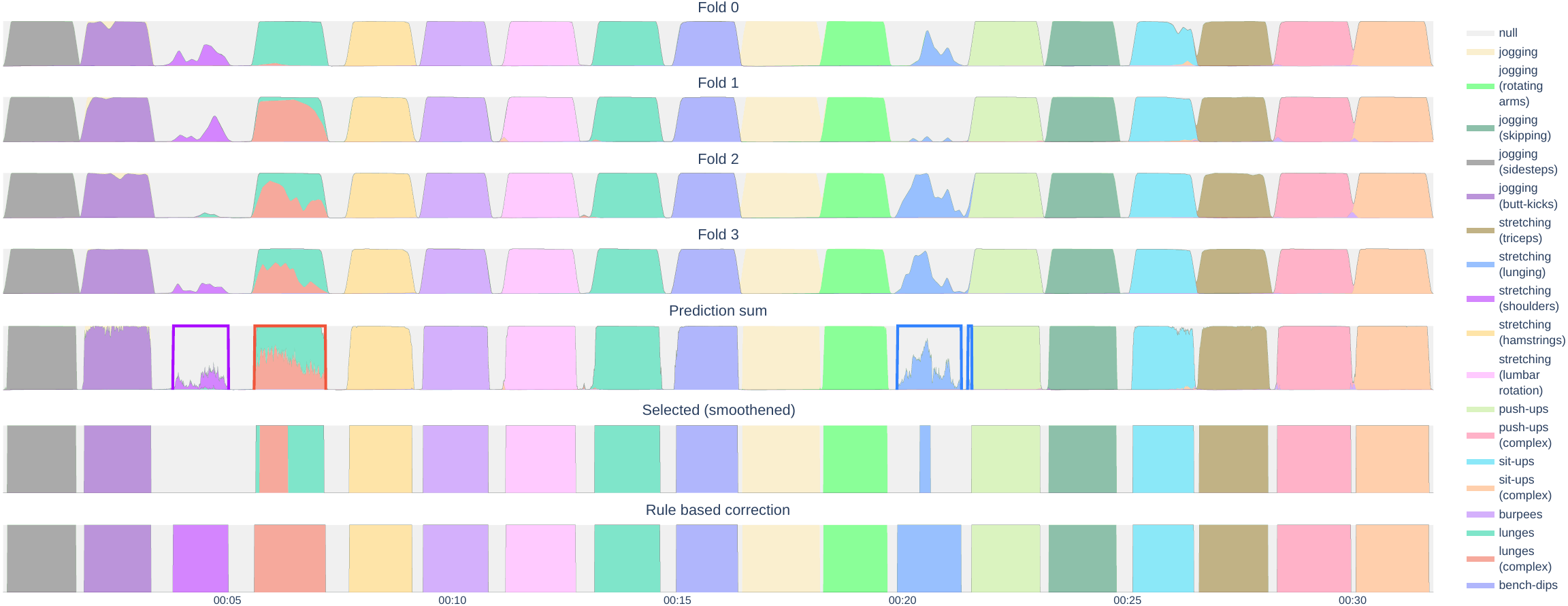}
  \vspace{-6mm}
  \caption{Stacked prediction plot of a test participant recording. The upper three subplots show the 0.5-second predictions for each of the three training folds. The fourth subplot displays the sum of these predictions, and the resulting k-fold majority-voted prediction is shown in the fifth subplot. The blue line in the "Prediction Sum" subplot illustrates our rule-based postprocessing, identifying regions where an underrepresented class (e.g., "stretching (lunging)") is present but not selected due to another class having a higher probability (e.g., "null"). The bottom subplot displays our final prediction, incorporating these rule-based postprocessing corrections applied to the smoothed majority-voted selection from the above plot.}
  \Description{Six vertically stacked time series plots indicate the per-fold predictions on a test user (plot 1-3), the sum of these predictions (plot 4), the selected prediction (plot 5), and the rule-corrected prediction (plot 6).}
  \label{fig:rule_based}
\end{figure*}

An exploratory analysis of the training set labels revealed that each of the 18 workout activities were present for at least 50 seconds for every participant recording. As such, we adopted this $>$50s presence rule as a post-hoc method to boost underrepresented classes in regions where other activities were selected (with low probabilities). Figure~\ref{fig:rule_based} demonstrates this boosting method applied to a user session from the test set.

\section{Experimental Results and Discussion}\label{sec:exp_results}

\begin{table}[t]
\caption{Macro F1 scores of the investigated approaches.}
\vspace{-3mm}
\label{tab:results}
\resizebox{\columnwidth}{!}{%
\begin{tabular}{l|cc|lc|c}
\toprule
 & \multicolumn{2}{c}{\textbf{Val. score}} & \multicolumn{2}{c}{\textbf{Data size}} &  \\
\textbf{Approach} & \textbf{F1} & \textbf{F1$_{PP}$} & \textbf{\# feats} & \textbf{$\mathbf{N_{augm}/N}$} & \textbf{rot. inv.} \\
\midrule
UL-pairing & 0.9154 & 0.9187 & 996 & 4 & $\pm$ \\
LR-stacking & 0.9084 & 0.9130 & 1992  & 4 & $\pm$  \\
raw & 0.8953 & 0.9001 & 1992 & 1 & $\times$ \\
rot\_inv$_{stat3}$ & 0.8837 & 0.8904 & 1992 & 1 & \checkmark \\
rot\_inv$_{sort}$ & 0.8815 & 0.8891 & 1992 & 1 &  \checkmark \\
rot\_inv$_{stat2}$ & 0.8806 & 0.8882 & 1328 & 1 &  \checkmark \\
SMV & 0.8055 & 0.8146 & 664 & 1 & \checkmark \\
\bottomrule
\end{tabular}
}
\vspace{-6mm}
\end{table}

In accordance with the WEAR dataset benchmarking~\cite{bock2023wear}, all experiments were conducted using grouped 3-fold cross-validation, with subjects being grouped per fold. 
Using the macro F1 metric, we analyze the predictions on the validation (out-of-fold) set and assess the impact of smoothing, indicated by respectively the $F1$ and $F1_{PP}$ columns in Table~\ref{tab:results}.
The experiments were performed on a server computer (Arch Linux), with an AMD Ryzen 5 2600x CPU, 48GB of DDR4 RAM, and an Nvidia RTX 2070 GPU. The total execution time for all experiments, including preprocessing, feature extraction, data augmentation, and postprocessing, was less than 1 hour.

A first notable observation from Table~\ref{tab:results} is that features derived from the \texttt{SMV} signal resulted in the poorest validation performance. Yet, \texttt{SMV} has been commonly employed in many wearable-based activity detection studies~\cite{wang2018benchmarking}.

Second, our proposed rotation-invariant statistical aggregation approaches (i.e. \texttt{rot\_inv$_{sort}$}, \texttt{rot\_inv$_{stat3}$}, and \texttt{rot\_inv$_{stat2}$}) outperformed the \texttt{SMV} configuration by an absolute margin of $\pm$7.5\%.
Interestingly, rot\_inv$_{stat2}$, which has a 1/3 compression ratio, showed only a minimal decrease in F1 score (absolute 0.1-0.3\%) compared to the other two, non-compressed, approaches. 

Third, the three rotation-invariant statistical aggregation techniques were surpassed by the \texttt{raw} approach by an absolute $\pm$1\%. This suggests that rotation information, captured by the axial components in the raw features, is of large importance for this challenge. Although our exploratory data analysis revealed that some recordings/sessions involved wearables being swapped or worn in an opposite orientation, the majority of the data demonstrated a consistent orientation. We suspect that this consistency limited the performance improvement from rotation-invariant feature aggregation. This insight led us to designing the left-right swapping and upper-lower limb pairing techniques, which retain orientation information while augmenting the data to enhance model robustness.

Fourth, both proposed data augmentation techniques outperformed the \texttt{raw} configuration, with an absolute improvement of $1\%$ for the LR-stacking and $1.5\%$ for the UL-pairing. Notably, UL-pairing achieved higher performance while utilizing features from only two wearables (per feature vector). This observation hints towards substantial redundancy between left-right wearable data. Moreover, the UL-pairing technique also proves robust to missing wearable data, as long as data from one of the left-right limb wearables is available.

Last, for all configurations, temporal prediction smoothing consistently improved performance, with absolute gains ranging from 0.3\% to 0.9\%.

Figure~\ref{fig:cm-3-fold} complements Table~\ref{tab:results} by showing the confusion matrices for our three best experiments. Note that these results can be positioned against the inertial baselines obtained by Bock et al.~\cite{bock2023wear}, since we used a similar grouped 3-fold CV procedure, the same stride (i.e., 0.5s), and temporal prediction smoothing. The highest macro F1 score from an inertial-only model in the WEAR dataset paper, achieved by the Attend and Discriminate (A- and D-) deep learning model architecture~\cite{abedin2021attend}, was 83.08\%. 
All our experiments, except for the SMV configuration, surpassed this performance by a substantial margin, demonstrating the expressiveness of a multi-resolution feature vector (using only limited, i.e., 14, feature functions).

Interestingly, all confusion matrices show limited confusion among
activity classes but substantial confusion with the null class. This is especially pronounced for stretching-based activities, which have the greatest margin for improvement. Additionally, we observe a notable improvement in distinguishing between lunges and complex lunges, with a 5\% absolute increase from the raw data to both data augmentation configurations.



\subsection{Test Set Predictions}
To generate predictions for the test set, we utilize the UL-pairing model along with all three postprocessing techniques detailed in Section~\ref{sec:postprocess}. Specifically, the UL-pairing model combined with smoothing achieved a macro F1 score of 91.87\% on the validation data (see Table~\ref{tab:results} and Figure~\ref{fig:cm-3-fold}).

Figure~\ref{fig:rule_based} illustrates the three postprocessing steps (k-fold majority voting, temporal prediction smoothing, and rule-based prediction boosting) applied to a participant recording from the test set. Since we employ a 0.5-second stride, predictions were made in 0.5-second intervals. To comply with the competition's requirement for sample-wise predictions, we expanded these 0.5-second predictions to the nearest sample-wise timestamp. Empirical validation of this expansion on the 3-fold results showed a negligible performance decrease of 0.02\% (resulting in an expanded macro F1 score of 91.85\%), confirming that a 0.5-second stride is adequately fine-grained.



\section{Conclusion}
This paper presents the findings and final approach of the "Signal Sleuths" team for the 2024 HASCA WEAR challenge. During exploratory data analysis, we identified inconsistencies in wearable orientation both within and across participant recordings, leading to the exploration of rotation-invariant techniques. 
We employed a fixed set of multi-resolution features and focused on a traditional machine learning pipeline, enabling us to conduct an ablation study on novel rotation-invariant and multi-wearable data augmentation techniques. Specifically, we investigated: (i) relying solely on the SMV features, (ii) a novel approach that makes rotation-aware features rotation-invariant through statistical aggregation, and (iii) two data augmentation techniques: swapping left-right wearables to enhance model robustness (LR-swapping) and using only one upper and one lower limb wearable pair (UL-pairing). In addition to these techniques, we evaluated the impact of temporal prediction smoothing as a postprocessing step.

Our results indicated that solely relying on SMV features yields the poorest performance. Yet, this has been commonly used as the de facto technique in many works.
Moreover, our proposed rotation-invariant statistical aggregation demonstrated a substantial improvement over SMV, but could not outperform the rotation-aware model (i.e. raw), underscoring the significance of rotation information in this dataset. Both LR-swapping and UL-pairing outperformed the raw model, with UL-pairing achieving the highest sample-wise macro F1 score of 91.85\% (using temporal prediction smoothing). Given that UL-pairing uses data from only one upper and one lower limb wearable, we hypothesize a high redundancy in left-right wearable data.
Furthermore, we assessed the impact of temporal prediction smoothing, which consistently improved performance by an absolute 0.3\% to 0.9\%.

Comparing our results with the benchmark approaches from the WEAR dataset paper, we demonstrated the competitiveness of a multi-resolution feature vector combined with a traditional machine-learning pipeline.

In conclusion, through our proposed multi-wearable data augmentation and several post-processing steps, we designed more robust models for multi-wearable workout activity detection.




\begin{acks}
Jonas Van Der Donckt (1S56322N) is funded by a doctoral fellowship of the Research Foundation Flanders (FWO).
\end{acks}

\bibliographystyle{ACM-Reference-Format}
\bibliography{references}


\end{document}